%% file: main.tex
\documentclass[sigconf]{acmart}
\usepackage{colortbl}
\usepackage{multirow}
\usepackage{makecell}
\usepackage{threeparttable}
\usepackage{arydshln}
\usepackage{soul}
\acmConference[KDD '27]{ACM SIGKDD Conference on Knowledge Discovery and Data Mining}{August 1--5, 2027}{San Jose, CA, USA}
\acmYear{2027}
\copyrightyear{2027}
\acmDOI{XXXXXXX.XXXXXXX}

\newcommand{\cmark}{\checkmark}

\usepackage{algorithm}
\usepackage{algorithmic}

\usepackage{array}
\usepackage{tabularx}

\newcolumntype{Y}{>{\centering\arraybackslash}X}

\newcommand{\best}[1]{{\bfseries\boldmath #1}}
\newcommand{\second}[1]{\underline{#1}}

\usepackage{makecell}

\usepackage{newfloat}
\usepackage{listings}
\DeclareCaptionStyle{ruled}{labelfont=normalfont,labelsep=colon,strut=off} 
\floatstyle{ruled}
\newfloat{listing}{tb}{lst}{}
\floatname{listing}{Listing}

\title{Towards AI-Driven Nanomedicine Discovery: A Benchmark and Multimodal Learning Framework for Nano Self-Assembly Prediction}

\author{Quan Hao}
\email{haoquan@emails.bjut.edu.cn}
\affiliation{%
  \institution{The School of Information Science and Technology, Beijing University of Technology}
  \city{Beijing}
  \country{China}}
  \authornotemark[1]

\author{Mengyue Fan}
\email{ fmengyue0927@163.com }
\affiliation{%
  \institution{Artemisinin Research Center, China Academy of Chinese Medical Sciences}
  \city{Beijing}
  \country{China}}
  \authornotemark[1]

\author{Zifan Dong}
\email{dongzifan@emails.bjut.edu.cn}
\affiliation{%
  \department{}
  \institution{College of Computer Science, Beijing University of Technology}
  \city{Beijing}
  \country{China}}

\author{Jianduo Zhao}
\email{zhao18210930562@163.com}
\affiliation{%
  \institution{Artemisinin Research Center, China Academy of Chinese Medical Sciences}
  \city{Beijing}
  \country{China}}

\author{Changhao Xiao}
\email{xiaochanghao@stu.gzy.edu.cn}
\affiliation{%
  \institution{Artemisinin Research Center, China Academy of Chinese Medical Sciences}
  \city{Beijing}
  \country{China}}

\author{Shangqing Jiao}
\email{Jiaoshangqing@emails.bjut.edu.cn}
\affiliation{%
  \institution{The School of Information Science and Technology, Beijing University of Technology}
  \city{Beijing}
  \country{China}}

\author{Hao Zhang}
\email{z5753441@ad.unsw.edu.au}
\affiliation{%
  \institution{School of Computer Science and Engineering, University of New South Wales}
  \city{Sydney}
  \country{Australia}}

\author{Yudong Wang}
\email{yudong.wang@ia.ac.cn}
\affiliation{%
  \institution{Institute of Automation, Chinese Academy of Sciences, University of Chinese Academy of Sciences}
  \city{Beijing}
  \country{China}}

\author{Fei Xia}
\email{fxia@icmm.ac.cn}
\affiliation{%
  \institution{Artemisinin Research Center, China Academy of Chinese Medical Sciences}
  \city{Beijing}
  \country{China}}

\author{Jigang Wang}
\email{jgwang@icmm.ac.cn}
\affiliation{%
  \institution{Artemisinin Research Center, China Academy of Chinese Medical Sciences}
  \city{Beijing}
  \country{China}}

\author{Liguo Zhang}
\email{zhangliguo@bjut.edu.cn}
\affiliation{%
  \institution{The School of Information Science and Technology, Beijing University of Technology}
  \city{Beijing}
  \country{China}}
  \authornotemark[2] 

\author{Chong Qiu}
\email{cqiu@icmm.ac.cn}
\affiliation{%
  \institution{Artemisinin Research Center, China Academy of Chinese Medical Sciences}
  \city{Beijing}
  \country{China}}
  \authornotemark[2] 

\renewcommand{\shortauthors}{Hao et al.}

\begin{document}

\setlength{\emergencystretch}{1em}

\begin{abstract}
Nano self-assembly organizes molecular components into bioactive nanoscale structures. Self-assembled nanoparticles (NAPs) derived from Chinese herbal formulas and applications such as anti-lung-cancer therapy demonstrate the substantial potential of self-assembly for nanomedicine discovery. Yet discovery still relies on costly wet-lab screening, while existing machine learning approaches lack standardized tasks, effective pairwise compatibility modeling, and public benchmarks with unified evaluation. To address these limitations, we formalize NSA prediction as a binary classification task for predicting self-assembly between molecular pairs and then establish NSA-Bench, the first public benchmark with curated molecular combinations, experimental conditions, self-assembly labels, and standardized evaluation protocols. We further develop NSA-Net, an interaction-aware multimodal framework that integrates complementary molecular evidence from graph topology, sequence semantics, and physicochemical descriptors to learn molecular-pair representations for self-assembly prediction. Extensive experiments on NSA-Bench show that NSA-Net achieves a ROC-AUC of $0.9470\pm0.0112$ (Small) and $0.9492\pm0.0062$ (Large). On the Small track, it surpasses the strongest machine-learning and graph-based baselines by 3.9 and 17.1 percentage points, respectively. Representation analyses reveal interpretable molecular characteristics associated with self-assembly prediction captured by the learned representations. Moreover, an NSA-Agent case study further demonstrates how NSA-Net predictions can support formulation refinement through experimental-condition-aware reasoning. Our code is available at https://github.com/developer-hq/NSA-Net.

\end{abstract}

\keywords{Nano Self-Assembly Prediction, Multimodal Molecular Learning, Benchmark Dataset}

\setlength{\emergencystretch}{2em}
\maketitle
\setlength{\emergencystretch}{1em}

\begin{figure}[bt]
    \centering
    \includegraphics[width=0.45\textwidth]{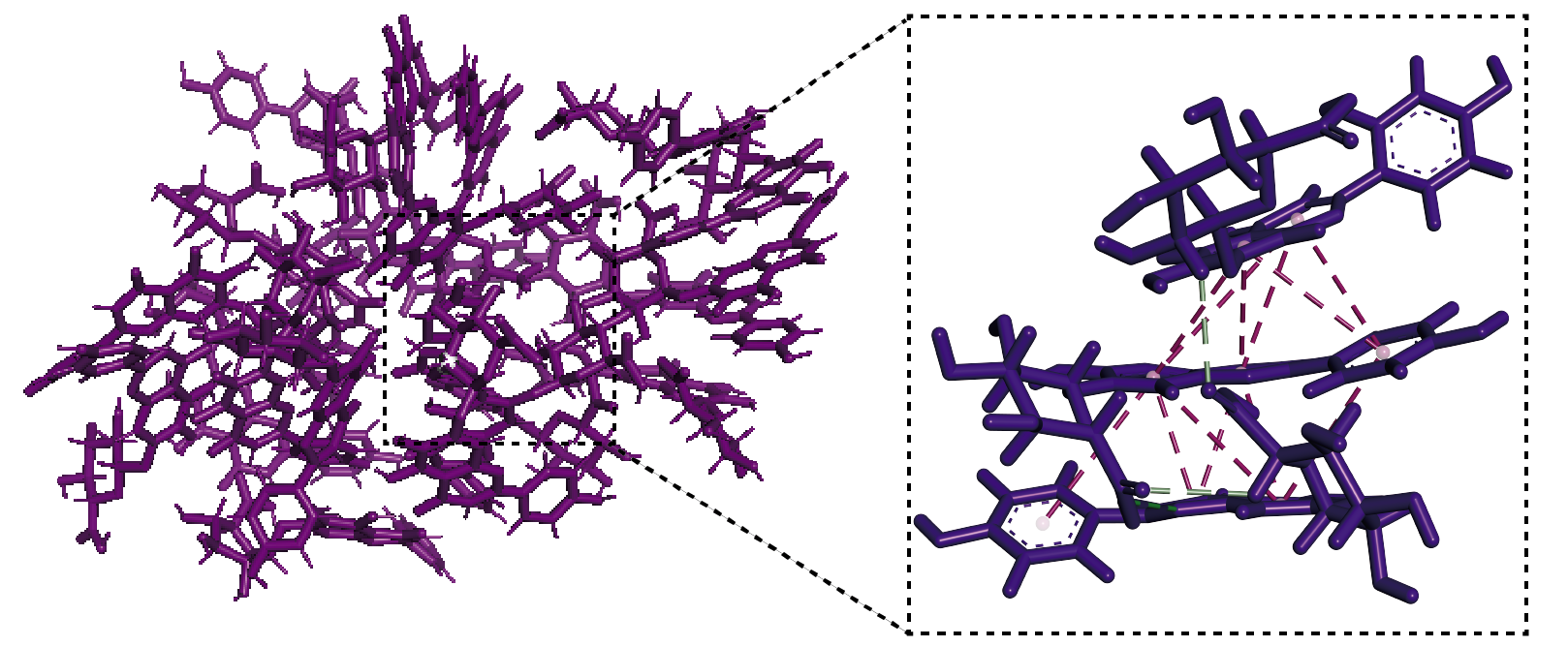}
\caption{Self-assembly-driven nanoparticle formation from a traditional Chinese medicine compound. Driven cooperatively by hydrogen bonding, hydrophobic interactions, and van der Waals forces, the compound molecules self-organize into nanoscale supramolecular architectures.}    \label{fig:illustration}
\end{figure}

\section{Introduction}

Nano self-assembly organizes molecular components into bioactive nanoscale structures and offers a promising route for new-drug development by controlling how molecular constituents are combined and presented \cite{reker2021selfassembling,zhang2025drugdrug}. In traditional Chinese medicine (TCM), Chinese herbal formula compatibility and aqueous decoction can generate supramolecular structures distinct from simple mixtures, and the resulting self-assembled nanoparticles constitute a material basis associated with bioactivity \cite{wang2025accessible}. The application of such nano-assembly particles derived from Chinese herbal formulas to anti-lung-cancer therapy further illustrates the value of nano self-assembly for drug development \cite{fu2025antilung}. Predicting whether molecular components will self-assemble under specified formulation conditions is therefore a scientifically and translationally important AI-for-Science task \cite{wang2026tcmsupramolecular,wang2025accessible}.

However, discovery remains dominated by costly wet-lab screening \cite{wang2024ionizable}. Expanding a molecular library combinatorially increases candidate pairs, while component ratios, concentration, pH, temperature, and incubation time multiply formulation choices per pair \cite{reker2021selfassembling,chan2025comet,chew2025mixtures}. Every combination requires preparation and observation, and the same pair may yield different outcomes across conditions \cite{bethiana2026mixing,chung2025rnaliposome,li2024nanocapsules}. This joint molecular–condition space makes exhaustive exploration impractical and reusable, standardized learning data scarce \cite{chew2025mixtures,chan2025comet,collins2026lnpdb}. Existing molecular AI extracts informative representations from molecular graphs, chemical strings, and physicochemical descriptors \cite{li2025finemoltex,ross2022molformer,moriwaki2018mordred,li2025min,chan2025comet}. However, as illustrated in Fig.~\ref{fig:illustration}, the self-assembly of traditional Chinese medicine compounds into nanoparticles is cooperatively driven by hydrogen bonding, hydrophobic interactions, and van der Waals forces—complex intermolecular interactions that remain difficult for current methods to capture adequately.

These gaps remain unresolved because dedicated computational work on nano-self-assembly prediction is still scarce. The absence of this shared infrastructure prevents experimentally informative records from supporting reproducible comparison, while the absence of explicit pair-level and context modeling leaves the central scientific relations underspecified \cite{zhang2025drugdrug,shan2026coassembly,qu2026foodderived,collins2026lnpdb}. To establish a common task and evaluation basis, we formalize nano-self-assembly prediction as binary classification over molecular pairs and introduce NSA-Bench. Through its publicly released NSA-Bench-Small protocol, NSA-Bench is, to our knowledge, the first public benchmark for this task, providing curated molecular pairs, formulation conditions, outcomes, provenance, fixed partitions, and unified metrics.

\begin{figure*}[ht]
    \centering
    \includegraphics[width=1\textwidth]{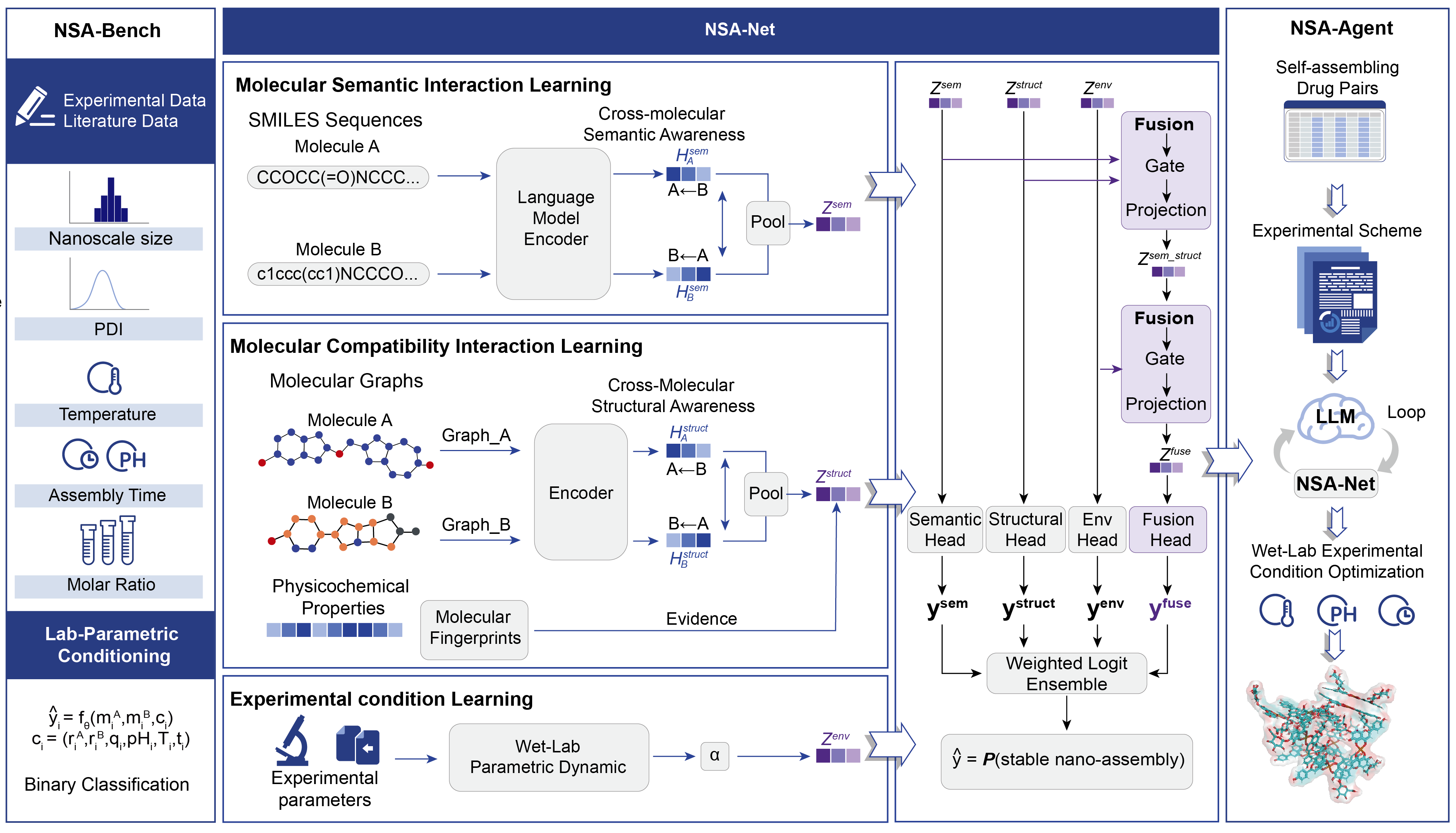}
    \caption{Overview of NSA-Bench, NSA-Net, and NSA-Agent, spanning benchmark construction, multimodal condition-aware prediction, and agent-guided experimental optimization.}
    \label{fig:framework}
\end{figure*}

To address the pairwise compatibility gap, we develop NSA-Net, an interaction-aware multimodal framework that learns a compatibility representation for each molecular pair. As illustrated in Fig.~\ref{fig:framework}, its primary structure-only configuration enables bidirectional information exchange between the two molecules in the SMILES and graph views, constructs order-invariant pair states, and complements them with physicochemical descriptors through GNN-anchored feature integration. A matched condition-aware extension further uses formulation variables to modulate cross-molecular graph interaction, allowing the same molecular pair to be interpreted differently across experimental settings. Thus, rather than encoding each molecule independently and combining the resulting embeddings only at prediction time, NSA-Net constructs multimodal evidence through the partner relation itself and, when conditions are available, qualifies that relation by the experimental context in which assembly is observed. In this way, NSA-Net addresses the pairwise compatibility gap by making partner-dependent molecular relations, rather than two isolated molecular summaries, the basis of prediction, while allowing experimental context to qualify the resulting compatibility estimate.

Extensive experiments on NSA-Bench show that NSA-Net achieves a ROC-AUC of $0.9470\pm0.0112$ (Small) and $0.9492\pm0.0062$ (Large). On the Small track, it surpasses the strongest machine-learning baseline (SVM) and graph-based baseline (Gramord) by 3.9 and 17.1 percentage points, respectively, while representation analyses identify interpretable molecular characteristics associated with self-assembly prediction. As an experimental extension, we further conduct an NSA-Agent case study on a specific Baicalin--Geniposide pair. Starting from a failed wet-lab experiment, the Agent analyzes the conditions and uses NSA-Net-based scores to refine ratio--time proposals over five rounds of 64 candidates, contracting the search space into a compact high-probability region. Its wet-lab recommendations cover conditions associated with qualified self-assembly observations in the existing experiments, showing that our framework can effectively translate predictions into actionable experimental guidance for nano-self-assembly. The resulting contributions are:

\begin{itemize}
    \item We formalize NSA prediction as pairwise binary classification and establish NSA-Bench---to our knowledge, the first public benchmark for this task---with curated molecular pairs, experimental conditions, labels, and standardized evaluation.
\item We introduce NSA-Net, an interaction-aware multimodal framework that captures microscopic self-assembly patterns by integrating graph topology, sequence semantics, and physicochemical descriptors to learn molecular-pair representations.
\item Experiments establish predictive effectiveness and interpretable molecular associations, while an NSA-Agent case study demonstrates the practical value of condition-aware reasoning for formulation refinement and nano-self-assembly development.
\end{itemize}

\section{Related Work}

\subsection{Nano-Molecular Self-Assembly Prediction and Computational Modeling}

Nano self-assembly organizes molecular components into bioactive nanoscale structures with potential for nanomedicine development. In traditional Chinese medicine (TCM), Chinese herbal formulas contain bioactive compounds \cite{wang2025accessible} that may spontaneously form self-assembled nanoparticles (NAPs) \cite{zhu2026plantderived}, with potential benefits for drug solubility, stability, bioavailability, and therapeutic efficacy \cite{li2024tcmnanodrug,liu2025naturalnanomedicines}. Their application to anti-lung-cancer therapy illustrates this value \cite{fu2025antilung}. However, identifying compatible molecular combinations and the formulation conditions under which they assemble remains experimentally intensive, motivating computational approaches to self-assembly prediction \cite{reker2021selfassembling,zhang2025drugdrug}.

Early computational studies mainly adopted structure-based strategies from drug discovery, including molecular docking, similarity search, and pharmacophore matching \cite{kumar2015hierarchical}. Subsequent machine learning approaches relied on molecular fingerprints and handcrafted descriptors \cite{rogers2010ecfp,moriwaki2018mordred}. Qu et al. \cite{qu2026foodderived} combined similarity search, pharmacophore matching, and random forests for self-assembly screening, while Shan et al. \cite{shan2026coassembly} integrated graph convolutional networks \cite{kipf2017gcn} with Mordred descriptors \cite{moriwaki2018mordred}. However, these methods remain limited by manual feature design and insufficient modeling of molecular cooperation \cite{qu2026foodderived,shan2026coassembly,chen2025molecbionet}.

Recent molecular representation learning approaches, including GNNs and molecular language models such as ChemBERTa \cite{chithrananda2020chemberta,luo2024mvmol,li2025finemoltex}, have enabled more powerful molecular modeling. Nevertheless, nano-molecular self-assembly prediction remains an emerging task, lacking public benchmarks and unified task definitions \cite{yang2025peptide,collins2026lnpdb,zhang2025drugdrug,qu2026foodderived,shan2026coassembly}. Existing methods rarely exploit modern molecular representations or explicitly model the coupled effects of molecular compatibility, intermolecular cooperation, and formulation conditions \cite{azagury2023prediction,shan2026coassembly,chan2025comet}.

\subsection{Multimodal Molecular Interaction Modeling and LLM-based Chemical Explanation}

Multimodal learning integrates molecular sequences, graph structures, and physicochemical descriptors for biomedical prediction \cite{lu2024multimodal,luo2024mvmol,li2025finemoltex,chen2025molecbionet}, typically through concatenation, attention, or gating \cite{peng2024mgndti,wei2025dacmf}, including gating-based molecular interaction models \cite{hua2025mmdgdti}. Yet current methods insufficiently model partner-dependent molecular interactions jointly with the formulation conditions required for nano-molecular self-assembly \cite{shan2026coassembly,chan2025comet,collins2026lnpdb}. Large Language Models (LLMs) have also been explored for molecular design, chemical knowledge retrieval, and scientific assistance \cite{boiko2023autonomous,bran2024augmenting,bran2026chemical}, but molecular predictors typically provide numerical outputs with limited interpretable reasoning \cite{lavecchia2025explainable,mirza2025chemical}. Grounded LLM assistance for nano-molecular self-assembly remains underexplored and must connect molecular evidence, experimental conditions, and predictive outcomes \cite{ashyrmamatov2026survey,mirza2025chemical,huang2026biomni}.

\section{NSA-Bench}
\label{sec:nsa_bench}

NSA-Bench establishes nano-self-assembly as a standardized AI-for-Science prediction task grounded in formulation experiments. Each record links a molecular pair and its preparation conditions to dynamic light scattering measurements, a curated binary outcome, and source provenance. To our knowledge, NSA-Bench is the first public benchmark for this task, with standardized records, fixed partitions, and unified evaluation.

\subsection{Task Definition}

For the evaluated two-component setting, observation $i$ is represented as
\begin{equation}
\mathcal{D}_i=(m_i^A,m_i^B,c_i,y_i),
\qquad
c_i=(r_i^A,r_i^B,q_i,\mathrm{pH}_i,T_i,t_i),
\end{equation}
where $m_i^A$ and $m_i^B$ denote the molecular components; $r_i^A$ and $r_i^B$ are their ratio entries; $q_i$, $\mathrm{pH}_i$, $T_i$, and $t_i$ denote total concentration, pH, temperature, and reaction or incubation time; and $y_i\in\{0,1\}$ is the curated experimental outcome. The standardized condition-aware task is
\begin{equation}
(m_i^A,m_i^B,c_i)\longmapsto y_i.
\end{equation}
A positive target denotes observed assembly under the recorded setting, whereas a negative target denotes an unsuccessful formulation. Because pair identity is order-independent, $(m^A,m^B)$ and $(m^B,m^A)$ form the same unordered group while retaining component-aligned ratios. The benchmark supports two evaluation settings:
\begin{align}
P(y=1\mid m^A,m^B)
& \quad \text{(structure-only)},\\
P(y=1\mid m^A,m^B,c)
& \quad \text{(condition-aware)}.
\end{align}
The two settings respectively evaluate molecular compatibility alone and compatibility qualified by experimental context, while remaining representation-agnostic.

\subsection{Benchmark Construction and Standardization}

\begin{figure*}[ht]
  \centering

\includegraphics[width=1\textwidth]{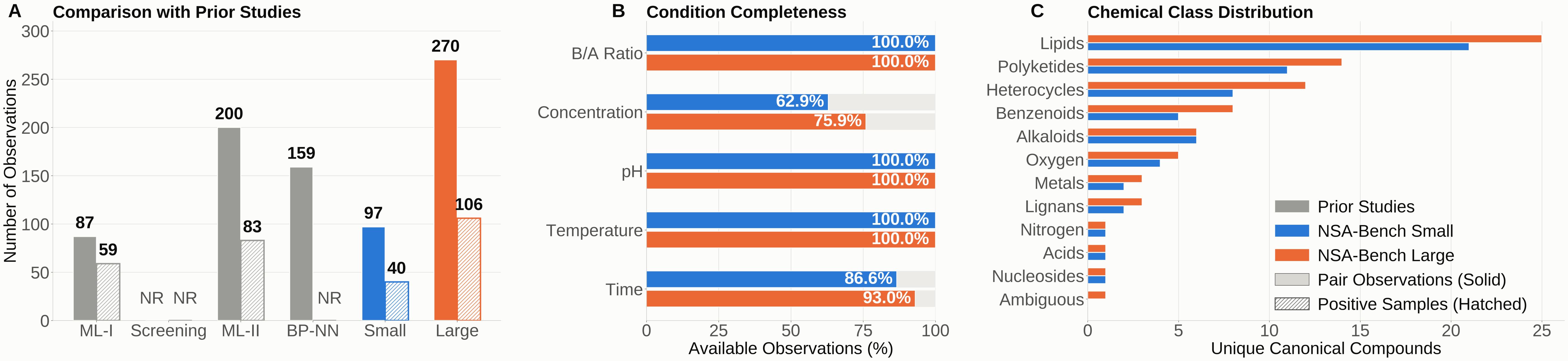}
   \caption{Characterization of NSA-Bench. (A) Comparison of NSA-Bench with prior nano-self-assembly studies interms of molecular-pair observations and positive samples; \textbf{NR} denotesnot reported. (B) Completeness of the experimental-condition annotations in NSA-Bench-Small and NSA-Bench-Large. (C) Chemical diversity of NSA-Bench}
  \label{fig:nsa_bench_profile}
\end{figure*}

NSA-Bench integrates traceable literature records with measurements from our wet-lab program and provides two tracks: NSA-Bench-Small contains 97 observations with 40 positive outcomes, while NSA-Bench-Large contains 270 observations with 106 positive outcomes. Compared with prior collections, which contain at most 200 observations and omit reusable data and experimental conditions, NSA-Bench offers greater scale and richer experimental context (Tab.~\ref{tab:nsa_bench_comparison}; Fig.~\ref{fig:nsa_bench_profile}A). Its extensive condition annotations include component ratio, concentration, pH, temperature, and time (Fig.~\ref{fig:nsa_bench_profile}B), while its compounds span multiple ClassyFire ChemOnt\cite{DjoumbouFeunang2016} superclasses (Fig.~\ref{fig:nsa_bench_profile}C). These properties support condition-qualified prediction that prioritizes both molecular pairs and formulation-space regions, reducing otherwise costly trial-and-error in wet-lab experimentation.

All records are standardized by molecular identity, SMILES, formulation conditions, binary outcome, and provenance. NSA-Bench-Small comprises 47 training observations (20 positive) and 50 held-out test observations (20 positive); NSA-Bench-Large comprises 220 training observations (86 positive) and 50 held-out test observations (20 positive). To control leakage, records sharing a source DOI or the same unordered standardized-SMILES pair are assigned to one partition, preventing direct source- and pair-level overlap between training and test data. ROC-AUC and AP evaluate discrimination and positive-class retrieval from continuous scores, while Precision and F1 evaluate thresholded prediction quality. NSA-Bench-Small is the released track; NSA-Bench-Large remains under restricted research access with delayed release planned.

\begin{table*}[ht]
\centering
\caption{Comparison of NSA-Bench with existing nano-self-assembly studies and datasets.}
\label{tab:nsa_bench_comparison}
\footnotesize
\renewcommand{\arraystretch}{1.18}
\setlength{\tabcolsep}{4pt}
\begin{tabularx}{\textwidth}{
    >{\raggedright\arraybackslash}X
    >{\raggedright\arraybackslash}p{3.3cm}
    >{\centering\arraybackslash}p{0.8cm}
    >{\centering\arraybackslash}p{1cm}
    >{\centering\arraybackslash}p{1.15cm}
    >{\centering\arraybackslash}p{1.05cm}}
\toprule
\textbf{Article Title}
& \textbf{Method}
& \textbf{Pairs}
& \makecell{\textbf{Public}\\\textbf{Data}}
& \makecell{\textbf{Samples}\\\textbf{Pos./Neg.}}
& \makecell{\textbf{Exp.}\\\textbf{Cond.}}
\\
\midrule
``Machine Learning-Driven Prediction, Preparation, and Evaluation of Functional Nanomedicines Via Drug--Drug Self-Assembly''
& Machine Learning
& 87
& NO
& 59/28
& NO
\\[2pt]
``A Computational Strategy for Identifying Self-Assembling Food-Derived Molecules for Antiparasitic Nanotherapy''
& Computational Screening
& NR
& NO
& NR
& NO
\\[2pt]
``AI-Assisted Engineering of Glycyrrhizic Acid/Simvastatin Nanocrystals for Multifunctional Treatment of Bacterial Osteomyelitis''
& Machine Learning
& 200
& NO
& 83/117
& NO
\\[2pt]
``An AI-Assisted Designed Supramolecularly Engineered Nanoplatform Reverses Pigmentation by Triggering an Ineffective Compensatory Melanin Production Program''
& Neural Network
& 159
& NO
& NR
& NO
\\
\midrule
\rowcolor{gray!18}
\textbf{NSA-Bench-Small}
& \textbf{Machine Learning}
& \textbf{97}
& \textbf{YES}
& \textbf{40/57}
& \textbf{YES}
\\
\rowcolor{gray!18}
\textbf{NSA-Bench-Large}
& \textbf{Machine Learning}
& \textbf{270}
& \textbf{YES}
& \textbf{106/164}
& \textbf{YES}
\\
\bottomrule
\end{tabularx}
\end{table*}

\section{Method}

As illustrated by the self-assembly of a traditional Chinese medicine compound in Fig.~\ref{fig:illustration}, nano self-assembly is cooperatively driven by hydrogen bonding, hydrophobic interactions, and van der Waals forces—microscopic intermolecular forces that existing methods struggle to capture. To address this, NSA-Net models nano self-assembly through partner-dependent molecular semantics, topology--physicochemical compatibility, and wet-lab conditions. Fig.~\ref{fig:framework} organizes these signals into semantic interaction, compatibility interaction, condition learning, and multimodal fusion. Each branch yields a pair representation and a perception signal, which are aggregated to predict the assembly outcome.

\subsection{Molecular Semantic Interaction Learning}

The predictive meaning of a molecular motif depends on its counterpart. Given semantic states $H_A^{\mathrm{sem}}$ and $H_B^{\mathrm{sem}}$ from a shared molecular-sequence encoder, the interaction operator $\mathcal I_{\mathrm{sem}}$ performs bidirectional information exchange as depicted in Fig.~\ref{fig:framework}, producing counterpart-aware states $\widetilde H_A^{\mathrm{sem}}$ and $\widetilde H_B^{\mathrm{sem}}$.

\begin{equation}
\label{eq:symmetric_pair}
\begin{aligned}
\phi(a,b)
&=\left[a+b,\ \lvert a-b\rvert,\ a\odot b\right],\\
Z^{\mathrm{sem}}
&=P_{\mathrm{sem}}\!\left(
\phi(u_A^{\mathrm{sem}},u_B^{\mathrm{sem}})
\right).
\end{aligned}
\end{equation}
The three terms encode shared activation, complementary variation, and feature co-activation. Since $\phi(a,b)=\phi(b,a)$, $Z^{\mathrm{sem}}$ is invariant to the input order. The Semantic Perception Head maps this state to semantic decision evidence.

\subsection{Molecular Compatibility Interaction Learning}

Assembly compatibility reflects both partner-aware structural organization and global physicochemical properties. Given structural states $H_A^{\mathrm{struct}}$ and $H_B^{\mathrm{struct}}$ from a shared molecular-graph encoder, $\mathcal I_{\mathrm{struct}}$ performs bidirectional information exchange as depicted in Fig.~\ref{fig:framework}, producing counterpart-aware states $\widetilde H_A^{\mathrm{struct}}$ and $\widetilde H_B^{\mathrm{struct}}$. Their readouts $u_k^{\mathrm{struct}}=\operatorname{Pool}(\widetilde H_k^{\mathrm{struct}})$ are combined by the same symmetric relation as $Z^{\mathrm{top}}=P_{\mathrm{struct}}(\phi(u_A^{\mathrm{struct}},u_B^{\mathrm{struct}}))$. The resulting state summarizes cross-molecular atom--bond organization.

Each molecule is also represented by a standardized Mordred descriptor vector $d_k$, encoded as $h_k^{\mathrm{pc}}$. Its symmetric pair relation enriches the topology state through
\begin{equation}
\label{eq:compatibility_state}
Z^{\mathrm{struct}}
=\operatorname{LN}\!\left[
Z^{\mathrm{top}}
+\gamma P_{\mathrm{pc}}\!\left(
\phi(h_A^{\mathrm{pc}},h_B^{\mathrm{pc}})
\right)
\right],
\end{equation}
where $P_{\mathrm{pc}}$ projects physicochemical evidence into the structural space and the learned scalar $\gamma$ controls its contribution. This residual retains topology as the identity path while incorporating pair-level evidence related to size, polarity, and hydrogen-bond capacity. The Structural Perception Head maps $Z^{\mathrm{struct}}$ to compatibility evidence.

\subsection{Experimental Condition Learning}

The assembly outcome of the same molecular pair can change with its formulation. The condition module encodes the component-aligned ratios, total concentration, pH, temperature, and reaction or incubation time, written as $c=(r_A,r_B,q,\mathrm{pH},T,t)$. A learned mapping $\alpha$ transforms these six variables into the condition state $Z^{\mathrm{env}}$, preserving their joint formulation pattern in a common latent space. The Condition Perception Head extracts environment-sensitive evidence, while condition-aware fusion uses this state to qualify the molecular-pair representation. Consequently, identical molecular inputs can yield condition-specific evidence as the recorded wet-lab setting changes.

\subsection{Multimodal Fusion and Prediction}

NSA-Net applies the two Fusion blocks sequentially. For an anchor representation $x$ and incoming evidence $y$, the fusion operation is defined as
\begin{equation}
\label{eq:multistage_fusion}
\begin{aligned}
g&=\sigma\!\left(W_g[x\Vert y]+b_g\right),\\
\operatorname{Fuse}(x,y)
&=\operatorname{LN}\!\left(W_xx+g\odot W_yy\right),\\
Z^{\mathrm{sem\text{-}struct}}
&=\operatorname{Fuse}(Z^{\mathrm{struct}},Z^{\mathrm{sem}}),\\
Z^{\mathrm{fuse}}
&=\operatorname{Fuse}(Z^{\mathrm{sem\text{-}struct}},Z^{\mathrm{env}}).
\end{aligned}
\end{equation}
Here, $W_g$ learns the feature-wise gate, while $W_x$ and $W_y$ project the two inputs into a common space. The first application enriches structural--physicochemical compatibility with semantic evidence, and the second incorporates the recorded wet-lab setting into the final pair representation.

The Semantic, Structural, Condition, and Fusion Perception Heads next translate their respective representations into two-dimensional decision-evidence vectors, $\mathbf y^k=\mathcal P_k(Z^k)$ for $k\in\{\mathrm{sem},\mathrm{struct},\mathrm{env},\mathrm{fuse}\}$. Their normalized weights satisfy $\omega_k\geq0$ and $\sum_k\omega_k=1$. The Weighted Logit Ensemble and final assembly score are
\begin{equation}
\label{eq:perception_ensemble}
\boldsymbol\ell=\sum_k\omega_k\mathbf y^k,
\qquad
\hat p=\left[\operatorname{softmax}(\boldsymbol\ell)\right]_{y=1}.
\end{equation}
Here, $\omega_k$ controls the contribution of each perception signal and $\hat p$ is the positive assembly score.

\subsection{NSA-Agent Construction}

A promising molecular pair still spans a large formulation space. NSA-Agent automates this search through the LLM--NSA-Net loop in Fig.~\ref{fig:framework}. At round $t$, the Agent proposes a batch $\mathcal C_t=\{c_{t,j}\}_{j=1}^{64}$ of feasible conditions, evaluates all candidates with the fixed condition-aware scoring pipeline, and records scores $p_{t,j}=f_{\mathrm{NSA}}(c_{t,j})$. The accumulated condition--score history then guides the Agent to contract the next ratio--time region around higher-ranked candidates, producing a propose--score--contract loop rather than sequential manual trial-and-error.

Convergence requires both the candidate conditions and their scores to concentrate within a high-ranking region. Let $\bar c_t$ and $\bar p_t$ denote the batch means, and let $D$ normalize the explored condition ranges. We summarize convergence at round $t$ with
\begin{equation}
\label{eq:agent_convergence}
\begin{aligned}
\zeta_t={}&
\lambda_c\max_j
\left\|D^{-1}(c_{t,j}-\bar c_t)\right\|_2\\
&+\lambda_p
\sqrt{\frac{1}{64}\sum_j(p_{t,j}-\bar p_t)^2}\\
&+\lambda_s(1-\bar p_t),
\end{aligned}
\end{equation}
where $\lambda_c$, $\lambda_p$, and $\lambda_s$ are nonnegative coefficients for condition dispersion, score dispersion, and ranking quality, respectively, with $\lambda_c+\lambda_p+\lambda_s=1$. Thus, $\lambda_c,\lambda_p,\lambda_s\in[0,1]$; increasing one coefficient makes convergence more sensitive to its corresponding criterion. The search is regarded as converged when $\zeta_t$ falls below a protocol-defined tolerance, while a maximum-round budget bounds the exploration. The converged batch is returned as a focused set of conditions for wet-lab follow-up, with model scores used for relative prioritization.

\begin{table*}[ht]
\centering
\caption{Performance comparison of machine learning baselines, graph-based methods, and our NSA-Net on the NSA-Bench-Small and NSA-Bench-Large tracks. Avg. Rank is computed over the four reported metrics within each track; lower is better, and ties receive average ranks.}
\label{tab:main_comparison}
\begin{tabularx}{\textwidth}{ll*{5}{Y}}
\toprule
Track & Method & Avg. Rank $\downarrow$ & ROC-AUC $\uparrow$ & AP $\uparrow$ & Precision $\uparrow$ & F1 $\uparrow$ \\
\midrule
Small & RF & $5.50$ & $0.8790\pm0.0048$ & $0.8242\pm0.0055$ & $0.6522\pm0.0000$ & $0.6977\pm0.0000$ \\
Small & GBDT & $6.75$ & $0.7673\pm0.0095$ & $0.6826\pm0.0146$ & $0.6650\pm0.0434$ & $0.7396\pm0.0353$ \\
Small & KNN & $5.50$ & $0.8673\pm0.0230$ & $0.8110\pm0.0255$ & $0.8746\pm0.0410$ & $0.6236\pm0.0870$ \\
Small & MLP & $4.50$ & $0.8577\pm0.0338$ & $0.8209\pm0.0197$ & $0.8667\pm0.0000$ & \second{$0.7429\pm0.0000$} \\
Small & SVM & \second{$2.75$} & \second{$0.9083\pm0.0000$} & \second{$0.8861\pm0.0000$} & \best{$0.9077\pm0.0237$} & $0.7060\pm0.0397$ \\
Small & LR & $3.50$ & $0.8767\pm0.0000$ & $0.8466\pm0.0000$ & $0.8667\pm0.0000$ & \second{$0.7429\pm0.0000$} \\
Small & Gramord & $6.25$ & $0.7758\pm0.0151$ & $0.7754\pm0.0226$ & $0.8001\pm0.0221$ & $0.7106\pm0.0299$ \\
\rowcolor{gray!18}
Small & \textbf{NSA-Net (Ours)} & \best{$1.25$} & \best{$0.9470\pm0.0112$} & \best{$0.8925\pm0.0158$} & \second{$0.9014\pm0.0420$} & \best{$0.8522\pm0.0365$} \\
\midrule
Large & RF & $5.25$ & $0.8907\pm0.0025$ & $0.7800\pm0.0119$ & $0.6641\pm0.0057$ & $0.7606\pm0.0120$ \\
Large & GBDT & $6.50$ & $0.8723\pm0.0035$ & $0.7478\pm0.0040$ & $0.6667\pm0.0000$ & $0.6829\pm0.0000$ \\
Large & KNN & $6.50$ & $0.8525\pm0.0156$ & $0.7472\pm0.0295$ & $0.6818\pm0.0000$ & $0.7143\pm0.0000$ \\
Large & MLP & $2.75$ & \second{$0.9358\pm0.0116$} & $0.8621\pm0.0237$ & $0.7600\pm0.0524$ & $0.8222\pm0.0263$ \\
Large & SVM & $3.75$ & $0.8917\pm0.0000$ & $0.7862\pm0.0000$ & \second{$0.7637\pm0.0384$} & $0.7448\pm0.0140$ \\
Large & LR & \second{$2.50$} & $0.9333\pm0.0000$ & \best{$0.8804\pm0.0000$} & $0.7308\pm0.0000$ & \second{$0.8261\pm0.0000$} \\
Large & Gramord & $7.50$ & $0.7958\pm0.2003$ & $0.7178\pm0.2100$ & $0.6510\pm0.2391$ & $0.7207\pm0.1485$ \\
\rowcolor{gray!18}
Large & \textbf{NSA-Net (Ours)} & \best{$1.25$} & \best{$0.9492\pm0.0062$} & \second{$0.8775\pm0.0250$} & \best{$0.8416\pm0.0213$} & \best{$0.8697\pm0.0115$} \\
\bottomrule
\end{tabularx}
\end{table*}

\begin{table*}[ht]
\centering
\caption{Ablation study of NSA-Net on the NSA-Bench-Small and NSA-Bench-Large tracks, evaluating the contribution of molecular evidence and experimental conditions. Values are five-seed mean $\pm$ sample SD. Bold marks the best reported mean within each track and metric.}
\label{tab:component_ablation}
\begin{tabularx}{\textwidth}{lcccc*{4}{Y}}
\toprule
Track & LM & GNN & Mordred & Fusion & ROC-AUC $\uparrow$ & AP $\uparrow$ & Precision $\uparrow$ & F1 $\uparrow$ \\
\midrule
Small & \cmark & -- & -- & -- & $0.8233\pm0.0446$ & $0.7881\pm0.0331$ & \second{$0.8504\pm0.0439$} & $0.7780\pm0.0160$ \\
Small & -- & \cmark & -- & -- & $0.7693\pm0.0135$ & $0.6414\pm0.0568$ & $0.6751\pm0.0234$ & $0.6910\pm0.0490$ \\
Small & \cmark & \cmark & -- & equal & $0.8127\pm0.0325$ & $0.7849\pm0.0404$ & $0.8400\pm0.0149$ & \second{$0.7802\pm0.0208$} \\
Small & \cmark & \cmark & -- & anchored & $0.8640\pm0.0366$ & $0.8181\pm0.0269$ & $0.8372\pm0.0461$ & $0.7283\pm0.0449$ \\
Small & -- & \cmark & \cmark & -- & \second{$0.9283\pm0.0305$} & \second{$0.8815\pm0.0297$} & $0.8383\pm0.0265$ & $0.7783\pm0.0592$ \\
\rowcolor{gray!18}
Small & \cmark & \cmark & \cmark & anchored & \best{$0.9470\pm0.0112$} & \best{$0.8925\pm0.0158$} & \best{$0.9014\pm0.0420$} & \best{$0.8522\pm0.0365$} \\
\midrule
Large & \cmark & -- & -- & -- & $0.9180\pm0.0058$ & $0.8275\pm0.0301$ & $0.7403\pm0.0189$ & $0.7949\pm0.0464$ \\
Large & -- & \cmark & -- & -- & $0.9180\pm0.0185$ & $0.8325\pm0.0211$ & $0.7862\pm0.0254$ & $0.7975\pm0.0361$ \\
Large & \cmark & \cmark & -- & equal & $0.9187\pm0.0097$ & $0.8400\pm0.0230$ & $0.7320\pm0.0065$ & $0.7734\pm0.0158$ \\
Large & \cmark & \cmark & -- & anchored & $0.8972\pm0.0100$ & $0.8114\pm0.0184$ & $0.7332\pm0.0209$ & $0.7776\pm0.0514$ \\
Large & -- & \cmark & \cmark & -- & $0.9407\pm0.0141$ & $0.8749\pm0.0385$ & $0.8273\pm0.0203$ & $0.8667\pm0.0213$ \\
\rowcolor{gray!18}
Large & \cmark & \cmark & \cmark & anchored & \best{$0.9492\pm0.0062$} & \second{$0.8775\pm0.0250$} & \second{$0.8416\pm0.0213$} & \best{$0.8697\pm0.0115$} \\
\midrule
\rowcolor{gray!18}
Condition & \cmark & \cmark & \cmark & anchored & \second{$0.9470\pm0.0079$} & \best{$0.8821\pm0.0151$} & \best{$0.8491\pm0.0321$} & \second{$0.8684\pm0.0106$} \\
\bottomrule
\end{tabularx}
\end{table*}

\section{Experiments}

\subsection{Experimental Setup}

We report ROC-AUC, average precision (AP), Precision, and F1 on fixed, leakage-controlled splits. Each configuration is independently trained with five random seeds under an identical protocol and evaluated once after training, without test-based selection. Precision and F1 use a fixed 0.5 threshold; results are reported as mean $\pm$ sample standard deviation across seeds.

\subsection{Comparison with Baselines}
\label{sec:comparison}

Under matched structure-only inputs, NSA-Net achieves the best average rank (1.25) on both tracks (Tab.~\ref{tab:main_comparison}). On NSA-Bench-Small, it leads in ROC-AUC ($0.9470\pm0.0112$), AP, and F1, exceeding the strongest conventional and graph-based baselines in ROC-AUC by 3.87 and 17.12 percentage points. SVM has slightly higher Precision, but its markedly lower F1 indicates a conservative operating point that misses more positives.

On NSA-Bench-Large, NSA-Net leads in ROC-AUC, Precision, and F1, while LR has marginally higher AP ($0.8804$ versus $0.8775$), consistent with a stable linear ranking signal that does not yield the best thresholded decisions. Following its reported graph-encoding, Mordred, and late-fusion design, Gramord performs less consistently and shows substantial run-to-run variation, plausibly because independently encoded high-dimensional branches are difficult to fit with limited data. In contrast, NSA-Net constructs symmetric, partner-conditioned relations and integrates physicochemical complementarity at the pair level, yielding more consistent ranking and classification performance.

\subsection{Component Ablation and Condition Study}
\label{sec:ablation}

The ablation study examines semantic interaction, compatibility interaction, physicochemical enhancement, and fusion. Semantic-only and compatibility-only variants isolate the two interaction paths; their equal- and anchored-fusion variants test evidence integration; the compatibility-plus-physicochemical variant isolates descriptor enhancement; and the full structure-only NSA-Net combines all molecular evidence. Every variant follows the same 65-epoch, five-seed protocol.

Physicochemical enhancement produces the clearest cross-track gains. Relative to compatibility interaction alone, adding physicochemical evidence raises ROC‑AUC/AP by 0.1590/0.2401 on NSA‑Bench‑Small and 0.0227/0.0424 on NSA‑Bench‑Large. Adding it to anchored semantic–compatibility fusion yields further gains of 0.0830/0.0744 and 0.0520/0.0661, establishing pair‑level physicochemical complementarity as a strong enhancement to learned structural evidence.

Semantic interaction and anchored fusion provide additional value. On NSA‑Bench‑Small, the full structure‑only NSA‑Net improves over compatibility plus physicochemical evidence by 0.0187 ROC‑AUC, 0.0110 AP, and 0.0739 F1. On NSA‑Bench‑Large, the full structure‑only NSA‑Net further improves all metrics over compatibility plus physicochemical evidence, with gains of 0.0085 ROC‑AUC, 0.0026 AP, 0.0143 Precision, and 0.0030 F1, showing that semantic integration strengthens balanced classification. Equal and anchored fusion further emphasize complementary ranking and thresholded‑decision behavior across the two tracks.

\begin{figure*}[ht]
    \centering
    \includegraphics[width=2\columnwidth]{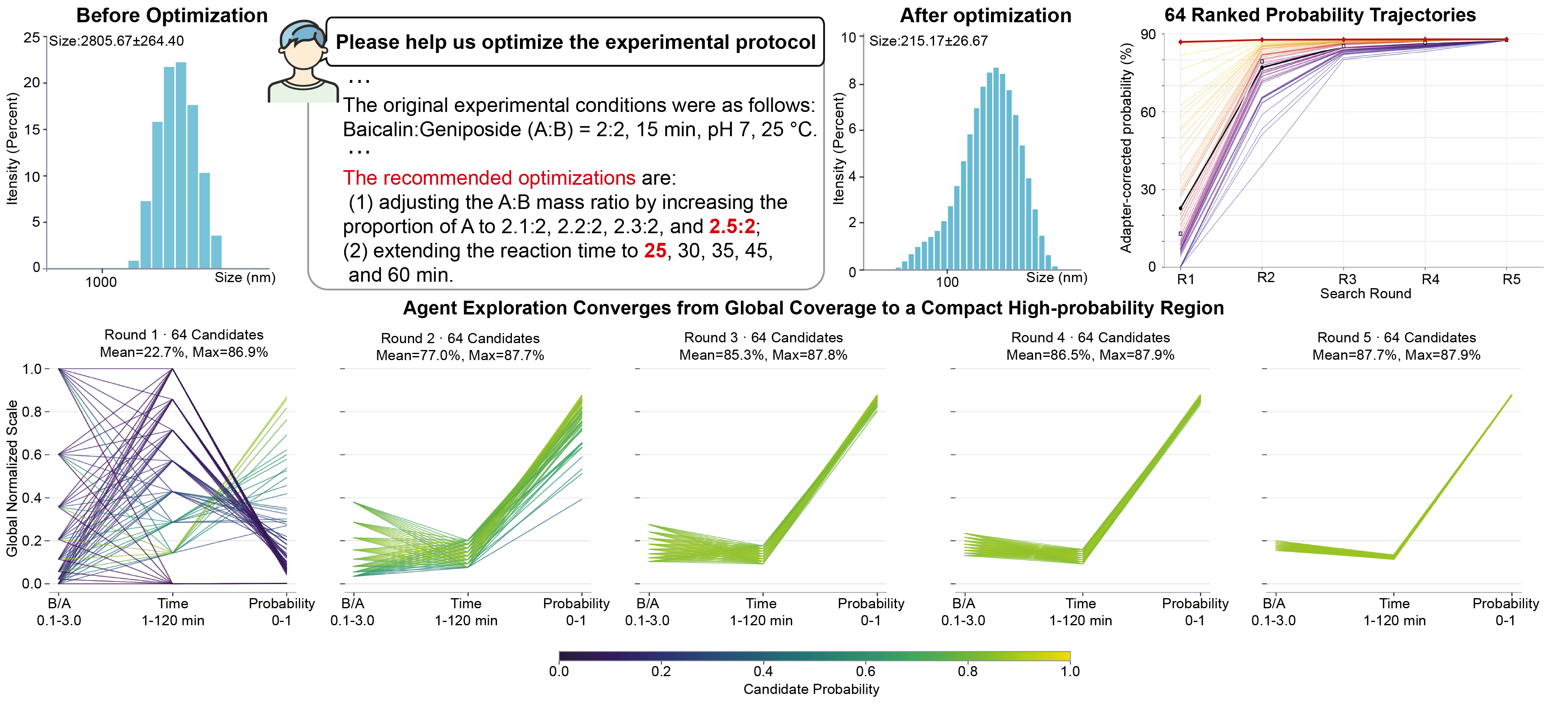}
    \caption{NSA-Agent uses NSA-Net-based qualification scores to refine the Baicalin--Geniposide formulation space. Five rounds of 64 ratio--time candidates progressively contract the search region and identify A:B $\approx2.209{:}2$ at 31.86 min as the highest-scoring proposal (83.12\%). The Agent also prioritizes the nearby $2.5{:}2$ setting represented in the wet-lab records. The upper DLS panels show the displayed particle-size distributions before and after experimental optimization ($2805.67\pm264.40$ and $215.17\pm26.67$ d.nm). Model scores guide candidate prioritization rather than represent wet-lab success rates, and the exact 31.86-min proposal remains to be tested.}
    \label{fig:agent_exploration}
\end{figure*}

\begin{figure}[ht]
    \centering
    \includegraphics[width=\columnwidth]{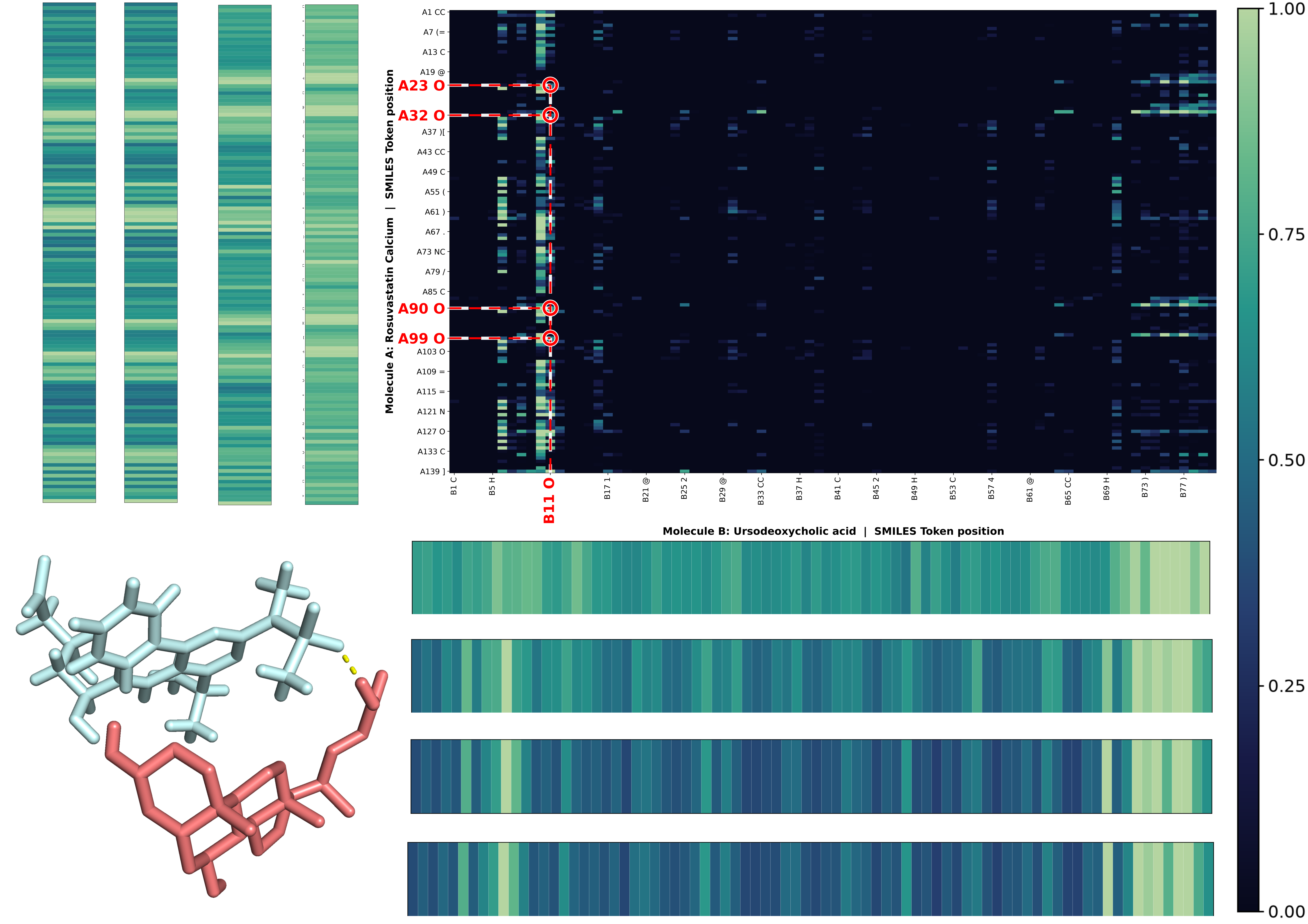}
    \caption{Feature analysis of top-ranked candidate interaction sites between Rosuvastatin Calcium and Ursodeoxycholic acid.}
    \label{fig:feature_analysis}
\end{figure}

\subsection{Representation and Feature Analysis}
\label{sec:feature_analysis}

To verify that NSA-Net captures the kind of microscopic interactions illustrated in Fig.~\ref{fig:illustration} (e.g., hydrogen bonding, hydrophobic interactions, and van der Waals forces), we examine the experimentally confirmed self-assembling pair Rosuvastatin Calcium (molecule A) and Ursodeoxycholic acid (molecule B)~\cite{Lee2024novel} through its learned representations. Fig.~\ref{fig:feature_analysis} unfolds the extracted features into pairwise heat sequences (molecule B on the x-axis, molecule A on the y-axis), where brighter cells contribute more strongly to the pair representation. Mapping these regions back to SMILES fragments reveals that NSA-Net highlights hydroxyl-containing regions in Rosuvastatin Calcium and the carboxyl-containing region in Ursodeoxycholic acid---a hydroxyl--carboxyl association pattern judged chemically plausible by domain experts. The heatmap further shows several distributed high-response regions rather than a single isolated coordinate, consistent with the many-to-many nature of nano self-assembly: multiple copies of the two molecules may adopt different relative arrangements, allowing several hydroxyl and carboxyl sites to participate. This partner-dependent perception demonstrates that NSA-Net jointly determines feature relevance across molecules and integrates multiple complementary regions into the pair representation. The correspondence between model-perceived regions and chemically plausible functional groups thus provides interpretable evidence that NSA-Net captures interaction characteristics relevant to self-assembly.

\subsection{Iterative Experimental Condition-Space Exploration}
\label{sec:agent_case}

We examine whether NSA-Agent can translate condition-aware predictions into focused formulation choices for the Baicalin--Geniposide pair. As shown in Fig.~\ref{fig:agent_exploration}, the case starts from an unsuccessful experiment at A:B $=2{:}2$, 15 min, pH 7, and $25\,^{\circ}\mathrm{C}$. Keeping molecular identity, pH, and temperature fixed, the Agent explores ratio and reaction time over five rounds of 64 candidates. The search contracts from broad coverage to a compact high-scoring region, while the mean candidate score rises from 22.7\% to 87.7\%. The ranked trajectories likewise converge across rounds, illustrating how NSA-Net scores can narrow the experimental search space.

The resulting suggestions increase A:B ratios to $2.1{:}2$--$2.5{:}2$ and extend reaction times to 25--60 min. These directions include A:B $=2.5{:}2$ at 25 min, a region associated with qualified observations in the existing records; the displayed DLS distribution also shifts from $2805.67\pm264.40$ to $215.17\pm26.67$ d.nm. Thus, NSA-Agent narrows a broad condition space into experimentally relevant regions for focused wet-lab follow-up.

\section{Conclusion}

This work establishes a reproducible foundation for AI-driven nano self-assembly. NSA-Bench standardizes formulation--condition observations and evaluation, with NSA-Bench-Small providing, to our knowledge, the first public benchmark for this task. NSA-Net learns interaction-aware molecular-pair representations and achieves a five-seed ROC-AUC of 0.9470, exceeding the strongest conventional and graph-based comparisons by 3.9 and 17.1 percentage points. Ablations support physicochemical complementarity and condition-qualified prediction. A retrospective NSA-Agent case further illustrates focused ratio--time prioritization consistent with existing wet-lab records, while its exact recommendation remains unmeasured. Together, these results provide a basis for prospective formulation testing.

\section{Limitations and Ethical Considerations}

NSA-Bench remains limited in scale and condition coverage, so estimates may be sensitive to sampling, provenance, and class balance. NSA-Agent provides retrospective, direction-level concordance with existing 25-min records not used at runtime; the exact recommendation and broader local ratio--time space require prospective testing. No human-subject or identifiable personal data are used. No identifiable personal data, human participants, or animal experiments were involved.

\section{Generative AI Usage}

Generative AI assisted with language polishing. The research design, analysis, interpretation, and final decisions were developed and verified by the authors, who take full responsibility for the content.

\clearpage

\bibliographystyle{ACM-Reference-Format}
\bibliography{aaai2026}

\appendix
\input{Appendix}
\end{document}

%% file: Appendix.tex
\label{app:bench}

\begin{table*}[ht]
\centering
\small
\setlength{\tabcolsep}{3pt}
\caption{Representative NSA-Bench records with verified data sources.}
\label{tab:app_doi_records}
\begin{tabularx}{\textwidth}{@{}
    >{\raggedright\arraybackslash}p{1.65cm}
    >{\hsize=0.9\hsize\raggedright\arraybackslash}X
    >{\hsize=1.1\hsize\raggedright\arraybackslash}X
    >{\centering\arraybackslash}p{0.9cm}
    >{\raggedleft\arraybackslash}p{0.75cm}
    >{\raggedleft\arraybackslash}p{0.72cm}
    >{\raggedleft\arraybackslash}p{0.8cm}
    >{\raggedleft\arraybackslash}p{0.5cm}
    >{\centering\arraybackslash}p{1.3cm}
    >{\centering\arraybackslash}p{0.85cm}@{}}
\toprule
ID & Molecule A & Molecule B & Ratio & Size & PDI & Conc. & pH & Temp./time & Labels \\
 & & & & (nm) & & (mM) & & ($^\circ$C/min) & \\
\midrule
self\_00177 & (s)-10-hydroxycamptothecin & chlorin e6 & 4:1 & 200.0 & 0.194 & -- & 7.4 & 25 / 9.5 & 1/1/1 \\
self\_00192 & Indocyanine Green & Paclitaxel & 1:5 & 176.0 & 0.331 & 0.5 & 7.4 & 25 / 60 & 1/1/1 \\
self\_00174 & oleanolic acid & chlorin e6 & 2.5:1 & 103.0 & 0.290 & -- & 7.0 & 25 / -- & 1/1/1 \\
self\_00139 & Camptothecine & Fe(III) porphyrin complex & 1:1 & 87.8 & 0.200 & 1.0 & 7.4 & 25 / 10 & 1/1/1 \\
self\_00210 & Gambogic acid & glycyrrhizic acid & 1:0.5 & 197.8 & -- & -- & 7.0 & 4 / 30 & 1/1/1 \\
self\_00220 & metformin & epigallocatechin gallate & 54.4:75 & 190.0 & -- & 0.5590 & 7.0 & 25 / 480 & 1/1/1 \\
self\_00130 & Berberine & oleanolic acid & 1:1 & 129.6 & 0.1723 & 1.0 & 7.5 & 60 / 30 & 1/1/1 \\
self\_00156 & quercetin & Poly(ethylene glycol) & 5:1 & 146.0 & 0.258 & -- & 7.4 & 25 / 5 & 1/1/1 \\
self\_00214 & Luteolin & 18-Beta-Glycyrrhetinic acid & 2:1 & 251.0 & 0.135 & -- & 7.4 & 25 / -- & 1/1/1 \\
self\_00209 & oleanolic acid & Evodiamine & 5:3 & 380.0 & 0.220 & -- & 7.0 & 2 / 480 & 0/1/0 \\
self\_00212 & Celastrol & Curcumin & 1:1 & 736.0 & -- & -- & 7.0 & 30 / 120 & 0/1/0 \\
self\_00164 & Berberine & (+)-catechin hydrate & 2:1 & 341.0 & 0.158 & -- & 7.4 & 60 / 60 & 0/1/0 \\
self\_00136 & 7-Ethyl-10-hydroxycamptothecin & Hydroxy Chloroquine & 1:5 & 55.0 & 1.150 & -- & 7.4 & 70 / 40 & 1/0/0 \\
\bottomrule
\end{tabularx}
\end{table*}

\begin{table*}[ht]
\centering
\small
\caption{NSA-Bench field dictionary. Component C is an optional source-schema field.}
\label{tab:app_dictionary}
\begin{tabularx}{\textwidth}{p{2.6cm}p{4.0cm}X}
\toprule
Field family & Source columns & Role and reporting rule \\
\midrule
Record identity & \texttt{self\_id} & Internal traceability identifier; provider names are excluded from manuscript tables. \\
Molecular identity & \texttt{chem\_A}, \texttt{chem\_B}, optional \texttt{chem\_C}; corresponding CAS fields & Links each component name to a chemical identifier. Optional C fields do not expand the evaluated A/B task. \\
Formulation ratio & A, B, and optional C ratio fields & Records the supplied component proportions for a formulation observation while maintaining alignment between each ratio and molecule. \\
Molecular structure & A, B, and optional C SMILES fields & Supplies structure strings for language and graph processing. Long SMILES are omitted from printed tables but remain in the data schema. \\
Formulation environment & Total concentration, pH, temperature, and reaction/incubation time & Supplies experimental Condition variables for the condition-aware model. These variables are distinct from Mordred molecular descriptors. \\
DLS outcomes & Particle size (nm) and PDI & Records observed size and dispersity measurements when available. \\
Provenance & Supplied DOI or empty publication field & Distinguishes DOI-backed literature records from in-house wet-lab observations. No DOI is inferred for an empty field. \\
Labels & \texttt{label\_nm}, \texttt{label\_pdi}, and final label & Stores curated size and PDI components and the final binary benchmark target. \\
\bottomrule
\end{tabularx}
\end{table*}

\begin{table*}[ht]
\centering
\small
\caption{DOI provenance corresponding to Table~\ref{tab:app_doi_records}.}
\label{tab:app_doi_sources}
\begin{tabularx}{\textwidth}{lXl}
\toprule
ID & Source DOI & Provenance type \\
\midrule
self\_00177 & \url{https://doi.org/10.2147/IJN.S224704} & DOI-backed \\
self\_00192 & \url{https://doi.org/10.1039/c9tb00687g} & DOI-backed \\
self\_00174 & \url{https://doi.org/10.1016/j.phymed.2021.153788} & DOI-backed \\
self\_00139 & \url{https://doi.org/10.1016/j.nantod.2022.101732} & DOI-backed \\
self\_00210 & \url{https://doi.org/10.1016/j.mtbio.2025.102278} & DOI-backed \\
self\_00220 & \url{https://doi.org/10.1016/j.mtbio.2025.102052} & DOI-backed \\
self\_00130 & \url{https://doi.org/10.1016/j.molstruc.2025.144577} & DOI-backed \\
self\_00156 & \url{https://doi.org/10.1016/j.cej.2023.142697} & DOI-backed \\
self\_00214 & \url{https://doi.org/10.1016/j.cclet.2025.111325} & DOI-backed \\
self\_00209 & \url{https://doi.org/10.1016/j.jddst.2025.107230} & DOI-backed \\
self\_00212 & \url{https://doi.org/10.1016/j.ijpharm.2025.125289} & DOI-backed \\
self\_00164 & \url{https://doi.org/10.1016/j.colsurfa.2024.134555} & DOI-backed \\
self\_00136 & \url{https://doi.org/10.1016/j.biomaterials.2022.121458} & DOI-backed \\
\bottomrule
\end{tabularx}
\end{table*}

\begin{table*}[t]
\centering
\small
\caption{Representative in-house wet-lab observations from the author-verified source table. No publication DOI is assigned.}
\label{tab:app_inhouse_records}
\begin{tabularx}{\textwidth}{@{}l>{\raggedright\arraybackslash}X>{\raggedright\arraybackslash}X*{7}{c}@{}}
\toprule
ID & Molecule A & Molecule B & Ratio & \makecell{Size\\(nm)} & PDI & \makecell{Conc.\\(mM)} & pH & \makecell{Temp./time\\($^\circ$C/min)} & Labels \\
\midrule
self\_00291 & Emodin & Shikonin & 1:3 & 158.9 & 0.133 & 20 & 7.0 & 25 / 3 & 1/0/0 \\
self\_00292 & Emodin & Shikonin & 1:4 & 325.9 & 0.119 & 20 & 7.0 & 25 / 3 & 0/0/0 \\
self\_00294 & Artesunate & baicalin & 2:1 & 9999 & -- & 30 & 7.4 & 40 / 3 & 0/1/0 \\
self\_00298 & Artesunate & Ginsenoside Rb1 & 1:1 & 335.3 & 0.603 & 20 & 7.4 & 40 / 3 & 0/0/0 \\
self\_00317 & Artesunate & Ginsenoside Rg1 & 1:2 & 116.7 & 0.407 & 30 & 7.4 & 40 / 3 & 1/0/0 \\
self\_00318 & Artesunate & Ginsenoside Rg1 & 1:2.5 & 115.7 & 0.444 & 35 & 7.4 & 40 / 3 & 1/0/0 \\
self\_00319 & Artesunate & Ginsenoside Rg3 & 2:1 & 9999 & -- & 30 & 7.4 & 40 / 3 & 0/1/0 \\
self\_00324 & Shikonin & Ginsenoside Rb1 & 1:1 & 9999 & -- & 20 & 7.4 & 40 / 3 & 0/1/0 \\
self\_00251 & Chlorogenic acid & Artemisinin & 1:0.25 & 3689 & 0.622 & 7.5 & 7.0 & 45 / 15 & 0/0/0 \\
self\_00006 & Artesunate & Puerarin & 1:1 & 7600 & 0.820 & 20 & 7.0 & 37 / 5 & 0/0/0 \\
self\_00036 & Chlorogenic Acid & Eupalinolide B & 0.6:0.25 & 256.0 & 0.220 & 20 & 7.0 & 25 / 5 & 1/1/1 \\
self\_00045 & gallic acid & paeonol & 1:2 & 261.7 & 0.394 & 20 & 7.0 & 40 / 10 & 1/1/1 \\
\bottomrule
\end{tabularx}
\end{table*}

\section{NSA-Bench Supplimentary Material}

\subsection{Overview of NSA-Bench}

Table~\ref{tab:app_doi_records} presents representative records from NSA-Bench with verified literature provenance. The selected entries span diverse molecular combinations, including natural products (e.g., oleanolic acid, berberine, quercetin), synthetic drugs (e.g., paclitaxel, metformin), and functional agents (e.g., chlorin e6, indocyanine green). Self-assembly outcomes are labeled with three-part annotations (DLS/UV-Vis/TEM), reflecting multi-evidence experimental validation rather than single-metric judgment. Successful assemblies (labels 1/1/1) typically exhibit particle sizes between 87.8 and 200.0 nm with PDI values below 0.3, indicating uniform nanoparticle formation. In contrast, records with at least one negative label (e.g., 0/1/0 or 1/0/0) show either excessively large sizes (341.0--736.0 nm) or high polydispersity (PDI up to 1.150), consistent with failed or poorly controlled assembly. The conditions further illustrate the practical formulation space: ratios range from 1:5 to 54.4:75, temperatures from 2 to 70$^\circ$C, and incubation times from 5 to 480 min, underscoring the combinatorial complexity that motivates condition-aware prediction. Together, these representative records demonstrate the diversity, standardization, and practical relevance of NSA-Bench as a benchmark for nano self-assembly prediction.

\subsection{Benchmark tracks and release boundary}

NSA-Bench represents each sample as a formulation--condition observation. A row records two molecular components, their aligned formulation ratios, available experimental conditions, dynamic light scattering outcomes, component labels, a curated final binary target, and source provenance. The same unordered molecular pair may occur in multiple rows when its ratio or environment changes.

NSA-Bench-Small is the current public protocol. Public47 contains 47 training records (20 positive and 27 negative), and the fixed Benchmark50 test set contains 50 records (20 positive and 30 negative), for 97 observations in total. Its split manifest, derived inputs, and multi-run result summaries are included in the public release. NSA-Bench-Large contains 220 training observations (86 positive and 134 negative) and a fixed 50-observation test protocol (20 positive and 30 negative); this track is planned for future public release to facilitate reproducibility and broader community benchmarking.

\begin{table*}[t]
\centering
\small
\caption{Supplied physicochemical profile used to formulate the NSA-Agent follow-up hypothesis.}
\label{tab:app_agent_properties}
\begin{tabularx}{\textwidth}{p{2.5cm}p{3.2cm}p{3.2cm}X}
\toprule
Property & Baicalin & Geniposide & Possible relevance to this case \\
\midrule
Molecular formula & $\mathrm{C}_{21}\mathrm{H}_{18}\mathrm{O}_{11}$ & $\mathrm{C}_{17}\mathrm{H}_{24}\mathrm{O}_{10}$ & Both are oxygen-rich glycosides with multiple potential hydrogen-bonding sites. \\
Molecular weight & 446.4 & 388.4 & Similar molecular weights allow a near-$1{:}1$ formulation without a large molar imbalance. \\
XLogP3 & 1.1 & $-2.3$ & Baicalin has greater hydrophobic/aromatic character, whereas Geniposide is more hydrophilic. \\
Topological polar surface area & $183\ \text{\AA}^{2}$ & $155\ \text{\AA}^{2}$ & Both molecules are highly polar, although their hydrophobic-core/hydrophilic-glycoside balances differ. \\
H-bond donors/acceptors & 6/11 & 5/10 & Both molecules can participate in multivalent hydrogen bonding and hydration-shell reorganization. \\
Structural feature & Planar flavone scaffold linked to glucuronic acid & Iridoid glycoside scaffold linked to glucose & Baicalin may provide an aromatic association surface, while Geniposide may contribute hydration and interfacial stabilization. \\
\bottomrule
\end{tabularx}
\end{table*}

\subsection{Training Settings and Training Dynamics}

NSA-Net is implemented in PyTorch and trained with AdamW. The batch size is 16, the weight decay is 0.07, and the maximum gradient norm is clipped to 10. For multi-seed experiments, we report the mean and standard deviation across different random seeds. We report ROC-AUC, AR, Precision, and F1.

\begin{figure}[htbp]
    \centering
    \includegraphics[width=0.8\columnwidth]{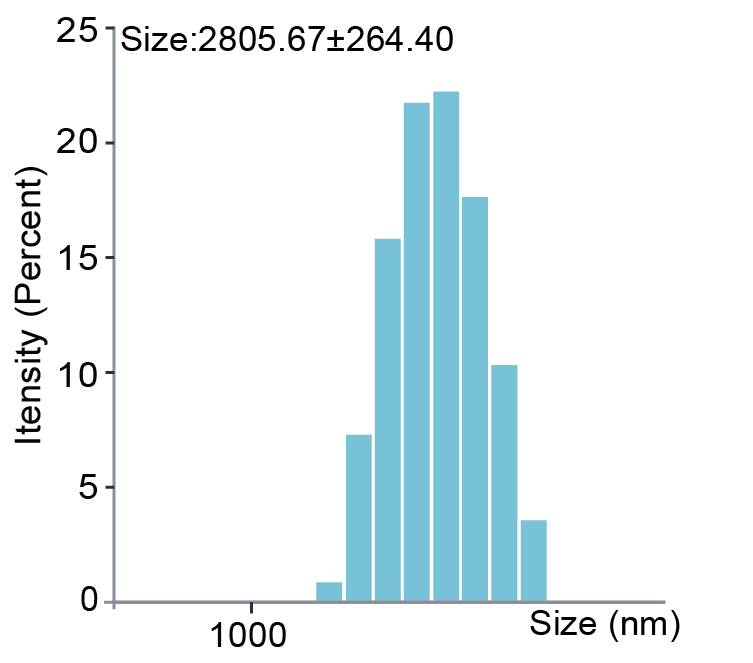}
    \caption{Intensity-weighted DLS particle-size distribution of the
    Baicalin--Geniposide formulation under the initial experimental
    condition. The measured hydrodynamic size is
    $2805.67\pm264.40$ d.nm, indicating a particle population dominated
    by micron-scale aggregates.}
    \label{fig:dls_initial}
\end{figure}

\begin{figure}[htbp]
    \centering
    \includegraphics[width=0.7\columnwidth]{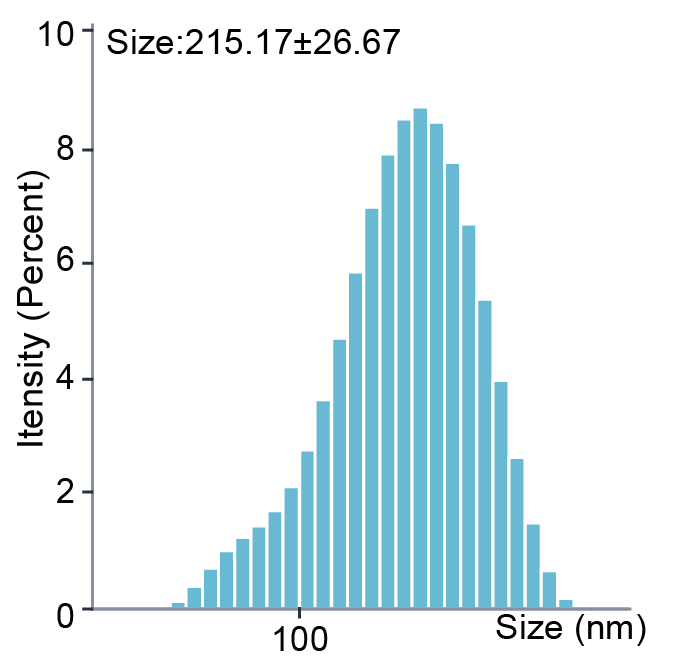}
    \caption{Intensity-weighted DLS particle-size distribution of the
    Baicalin--Geniposide formulation after experimental-condition
    refinement. The measured hydrodynamic size decreases to
    $215.17\pm26.67$ d.nm, indicating that the dominant particle
    population shifts from the micrometer scale to the nanometer scale.}
    \label{fig:dls_refined}
\end{figure}

\section{Reproducibility and Release Notes}
\label{app:release}

The public NSA-Bench-Small result package provides the split manifest, run-level predictions from five runs, and aggregated summary metrics. These artifacts document the evaluated sample assignment, expose run-to-run variation, and support recomputation of the reported results from the released prediction outputs. ROC-AUC and AP are computed directly from continuous scores, while threshold-dependent classification metrics use a fixed cutoff of 0.5. Results across the five runs are summarized as the mean with sample SD.

\section{Wet-Lab Results Appendix: Experimental Preparation and DLS Characterization}
\label{app:experimental_characterization}

Baicalin and Geniposide were separately dissolved in
phosphate-buffered saline (PBS) to obtain clear solutions or uniform
dispersions. The two drug solutions were then combined in a
round-bottom flask at the prescribed formulation ratio. The solvent
was removed under reduced pressure by rotary evaporation, allowing the
drug mixture to form a uniform thin film on the inner wall of the
flask. Fresh PBS was subsequently added to hydrate the film, followed
by ultrasonication to obtain the self-assembled nanoparticle
dispersion.

This film-hydration procedure was used as the common preparation
protocol for the Baicalin--Geniposide case study. Consistent with the
condition-aware prediction task defined in the main text, the molecular
identities and the general preparation procedure were kept unchanged
throughout the condition-space exploration. NSA-Agent varied only the
component-aligned A:B ratio and the recorded reaction or incubation
time, while pH and temperature were fixed at 7 and
$25\,^{\circ}\mathrm{C}$, respectively.

Therefore, NSA-Agent does not redesign the experimental preparation
procedure itself. Instead, it uses NSA-Net-based qualification scores
to rank candidate settings within a predefined and experimentally
feasible formulation space. Particle size and PDI are measured after
sample preparation and are used to characterize the resulting
dispersion and determine the experimental outcome. Neither particle
size nor PDI is supplied to NSA-Net or NSA-Agent as a runtime input.

The intensity-weighted dynamic light scattering (DLS) distributions
provide experimental characterization of the Baicalin--Geniposide
formulations before and after condition refinement. As shown in
Figure~\ref{fig:dls_initial}, the formulation prepared under the
initial condition exhibits an apparent hydrodynamic size of
$2805.67\pm264.40$ d.nm. This micron-scale particle population
indicates the presence of large aggregates or an insufficiently
dispersed formulation rather than a compact nanoscale dispersion.

After adjustment of the formulation conditions, the dominant
hydrodynamic size decreases to $215.17\pm26.67$ d.nm, as shown in
Figure~\ref{fig:dls_refined}. The substantial reduction in apparent
particle size is consistent with reduced coarse aggregation and the
formation of a smaller nanoparticle population. The dominant particle
population is therefore shifted from the micrometer scale to the
nanometer scale.

This comparison supports the condition-aware premise of NSA-Net and
NSA-Agent: even when the molecular pair and the general preparation
procedure remain unchanged, variations in component ratio and reaction
or incubation time can materially alter the physical state of the
resulting dispersion. The DLS observations provide experimental
support for prioritizing formulation conditions rather than treating
molecular compatibility as independent of the preparation environment.

Nevertheless, the DLS measurements do not independently establish the
specific molecular mechanism responsible for self-assembly. Because
intensity-weighted DLS is particularly sensitive to large particles,
the peak heights in Figures~\ref{fig:dls_initial}
and~\ref{fig:dls_refined} should not be directly compared, especially
because the two plots use different vertical-axis scales. In addition,
dispersion uniformity should be evaluated together with the
corresponding PDI rather than inferred solely from the apparent width
of the particle-size distribution. Complementary characterization and
prospective experiments are therefore required to validate the
intermolecular interactions and the exact condition selected by
NSA-Agent.